\documentclass[a4paper,12pt]{article}

\usepackage{jheppub} 

\usepackage[T1]{fontenc} 

\title{\boldmath Langlands Dualities through Bethe/Gauge Correspondence for 3d Gauge Theories}

\author[a]{Xiang-Mao Ding}
\author[b,1]{and Ting Zhang \note{Corresponding author.}}

\affiliation[a]{Institute of Applied Mathematics, Academy of Mathematics and Systems Science, \\
Chinese Academy of Sciences, Beijing 100190, China}
\affiliation[b]{School of Mathematical Sciences, Peking University, Beijing 100871, China}

\emailAdd{xmding@amss.ac.cn}
\emailAdd{zhangting@amss.ac.cn}

\abstract{
For non-simple laced Lie algebras, the $\text{B}_{N}$ and $\text{C}_{N}$ are Langlands dual to each other in mathematical. In this article, we give another Bethe/Gauge correspondence between 3d (or 2d) classical Lie group supersymmetry gauge theory with closed and open $\text{XXZ}$ (or $\text{XXX}$) spin chain. Here, the representations of the $\text{ADE}$ Lie algebras are self-dual, and while for the non-simple laced Lie algebras $\text{B}_{N}$ and $\text{C}_{N}$, their roles are exchanged in contrast with the results in \cite{DZ23a}. From Bethe/Gauge correspondence point of view, the two types of the effective superpotentials are Langlands duality to each other. For the $\text{B}_{N}$-type Lie algebra, a remarkable feature is that, to fix the spin sites by boundaries through Bethe/Gauge, the spins of the sites will be reversed. This is similarly to the so called electron-hole effect, we call this as a boundary-spin effect, a new kind of duality.
}


\begin{document} 
\maketitle
\flushbottom

\section{Introduction}
\label{sec:intro}

The integrability of 4d $\mathcal{N}=2$ supersymmetric gauge theories has aroused intense interest since the groundbreaking work of Seiberg and Witten, where the moduli space of vacua in 4d $\mathcal{N}=2$ gauge theory exhibit a fascinating correspondence
to the algebraic classical integrable systems \cite{GKMMM95,SW94a,SW94b,SW97}. In \cite{NS09a,NS09b}, Nekrasov and Shatashvili gave the relation between supersymmetry gauge theory and quantum integrable system, which was afterwards named as Bethe/Gauge correspondence. In this relationship, the effective twisted superpotential of a gauge theory corresponds to the Yang-Yang function of a quantum integrable system. With this regard, the space of supersymmetric vacua is equivalent to the state space of a quantum integrable system, whose Hamiltonian is the generator of the (twisted) chiral ring. That is to say, the spectrum of the quantum Hamiltonians coincides with the spectrum of the (twisted) chiral ring. The vacuum equation is
\begin{equation}\label{1}
    \text{exp}\left(\dfrac{\partial W_{\text{eff}}
    (\sigma)}{\partial \sigma_{i}}\right)=1
\end{equation}
where $W^{3d}_{\text{eff}}$ is called effective twisted superpotential, $\sigma$ is the eigenvalue of the complex scalar in the vector multiplet, $m$ is a twisted mass parameter. 
The Yang-Yang function $Y(u)$ gives the potential of an integrable system, and $u$ is the spectral parameter.

This algebraic classical correspondence is nicely lifted to the quantum correspondence by NS limit. In NS-limit, there are two $\Omega$-parameters $\epsilon_{1}$ and $\epsilon_{2}$. If $\epsilon_{1}=\epsilon_{2}=0$, the supersymmetry gauge theory corresponds to a classical algebra integrable system. If $\epsilon_{1}=h$ and $\epsilon_{2}=0$, the supersymmetry gauge theory corresponds to a quantum integrable system, where $h$ plays the role of Planck constant. The prepotential $\mathcal{F}(a,m;q)$ of the effective low-energy theory relates to the gauge theory partition $\mathcal{Z}$-function in the flat space limit as follows:
\begin{equation*}
\mathcal{F}(\mathfrak{a},m;q)=-\lim_{\epsilon_{1},\epsilon_{2}\rightarrow 0}\epsilon_{1}\epsilon_{2}\text{log}\mathcal{Z}(\mathfrak{a},m;q;\epsilon_{1},\epsilon_{2})
\end{equation*}
where $q$ here denotes the set of gauge coupling constants of the theory, $m$ the set of
masses of the hypermultiplets fields, and $\mathfrak{a}$ is the set of the flat special coordinates on the moduli space of vacua $\mathfrak{M}$ on the Coulomb branch of the theory. The geometry of $\mathfrak{M}$ and $\mathcal{F}(\mathfrak{a},m;q)$ is captured by a complex analogue of a classical integrable system, an algebraic integrable system. In terms of $\Omega$-backgrounds, the effective twisted superpotential is identified as 
$$\mathcal{W}(\mathfrak{a},m;q,\epsilon)=-\lim_{\epsilon_{2}\rightarrow 0}\epsilon_{2}\text{log}\mathcal{Z}(\mathfrak{a},m;q;\epsilon_{1}=h,\epsilon_{2})$$
The effective twisted superpotential is the same to (\ref{1}) \cite{LN21,NPS18,NS09c}.

From Lie theory, the $\text{ADE}$ series are self-dual, the $\text{B}_N$ and $\text{C}_N$ Langlands dual to each other. In mathematical, the role of the simple roots and  fundamental weights of a Lie algebra is exchanged. In physics, it is the electric-magnetic duality, or the GNO duality \cite{GNO77}. For a 4d $\mathcal{N}=1$ supersymmetric gauge theory, the Seiberg duality \cite{S95} is a version of electric-magnetic duality in supersymmetric gauge theory.  For supersymmetric QCD, it identifies in the infrared the quarks and gluons in a theory with $N_f$ quark flavors and $\text{SU}(N_c=N)$ gauge group for $N_f-N\ >1$ with solitons in a theory of $N_f$ quark flavors and $\text{SU}(N_f-N)$ gauge group. In a special case, $N_f= 2N$, with all the quiver nodes having a gauge group with the same rank $N$. One advantage of this choice is that the rank of the gauge group, and hence the number of the components of the spin of the integrable model at a lattice site, is preserved by the Seiberg duality. The 4d $\mathcal{N}=1$ Seiberg dualities will be inherited as the duality between two supersymmetric quiver gauge theories, as it be realized as the duality between different solutions of the Yang-Baxter equation \cite{YY15}, and the quiver theory can be obtained by glue three $R$-matrices through a product of three $R$-matrices.

In \cite{DZ23a}, we gave a uniform Bethe/Gauge correspondence between (2d or) 3d $\mathcal{N}=2$ supersymmetry gauge theory with classical Lie algebra and the ($\text{XXX}$ or) $\text{XXZ}$ spin chain with periodic boundary condition and open boundary condition, and we extend the new Bethe/Gauge correspondence between 3d quiver Lie algebra and high rank spin chain with different boundary condition \cite{DZ23b}. So we solved the problem left in \cite{KZ21} where $\text{C}$-type gauge theory could not match the open $\text{XXZ}$ spin chain obviously, and unified the correspondence by using the root structure of the Lie algebra. Different from the vacuum equation (\ref{1}), we used the modified vacuum equation
\begin{equation}\label{2}
    \text{exp}\left(\beta_{2}i \dfrac{\partial}{\partial \sigma}W^{3d}_{\text{eff}}(\sigma,m)\right)=1
\end{equation} 
where $\beta_{2}$ is the $U(1)$ charge fugacity, $i$ is the imaginary unit, and we found that the diagonal boundary conditions of the open $\text{XXZ}$ spin chain must take specific values to match the vacuum equations for different gauge groups, in there, the vacuum equations for the $B_N$ and $C_N$ belong to two different branches.

In this article, we present an alternate effective superpotential of the 3d supersymmetric gauge theory, and give another kind of Bethe/Gauge correspondence. Here, we raise the ratio of 4 and the root lengths of the Lie algebra to the power of the function instead of using it as a coefficient of the effective superpotential in\cite{DZ23a}. In the way, we set different Lie group representations, which is less anti-fundamental representation than \cite{DZ23a}. Alike the previous work, we also consider the two branches of the vacuum equations. Specially, we find that the boundary parameters of the open $\text{XXZ}$ spin chain which correspond to the $\text{B}$-type Lie algebra has magical phenomena, while the parameters matched the $\text{C}$-type Lie algebra look simply. For the $\text{A}$-type Lie algebra and the $\text{C}$-type Lie algebra, the boundary conditions are the same as the results in \cite{DZ23a}. 

That is to say, the boundary parameters of the open $\text{XXZ}$ spin chain which correspond to the $\text{B}$-type and the $\text{C}$-type Lie algebras look more naive and concise with \cite{DZ23a}, and $\text{B}$- and $\text{C}$-type gauge theories can be put into one branch vacuum equation with appropriate chosen boundary parameters. In other word, the realization given here just exchange the role of $\text{B}$- and $\text{C}$-type Lie algebra with in \cite{DZ23a}. As for the $\text{A}$-type and the $\text{D}$-type supersymmetry gauge theories, the corresponding boundary parameters are the same as the values in \cite{DZ23a}. From the Bethe/Gauge correspondence sense, we obtain another kind of effective superpotential realization, which is Langlands duality the ones in \cite{DZ23a}. 

This paper is organized as follows: in section \ref{a} we start with definition of the gauge theory partition function of 3d $\mathcal{N}=2$ on $S^{1}\times D^{2}$ and introduce the one-loop determinant of the vector multiplet and the chiral multiple, which are convenient to study the effective superpotential. In particular, we give a renewed effective superpotential using the property of Lie algebra root system. In section \ref{b}, we calculate the exact effective superpotential of the $\text{BCD}$-type gauge theory using the new representation we set. Next, in section \ref{c}, we give the correspondence between classical Lie group gauge theory and spin chains, including the spins of particle and boundary parameters. Then, in the next section, we compare the two different effective superpotentials. We try to explain the difference between the two-types boundary parameters of the open $\text{XXZ}$ spin chain related to the classical Lie group supersymmetry gauge theory, especially for $\text{B}$-types and $\text{C}$-type Lie algebra. In section \ref{e}, we calculate the effective superpotential and corresponding vacuum equations of the exception Lie algebras $E_{6,7,8}$ and $F_{4}$. In section \ref{f}, we give a conclusion and discussion. We give a brief recall of the integrable spin chains and the Bethe ansatz equation in a separate appendix \ref{A}.

\section{The gauge theory}\label{a}

\subsection{The one-loop determinant}

Now we give a brief review of the relevant gauge theories. In \cite{YS20}, the disk partition function $\mathcal{I}$ of 3d $\mathcal{N}= 2$ theory on $D^{2}\times S^{1}$ and 2d $\mathcal{N} = (2,2)$ theory on $D^{2}$ have been given,
\begin{equation*}\label{}
    \mathcal{I}=\dfrac{1}{|W_{G}|}\int \dfrac{d^{N}}{(2\pi)^{N}}e^{-S_{cl}}Z_{\text{vec}}Z_{\text{chi}}Z_{\text{bd}},
\end{equation*}
where $W_{G}$ is the Weyl group of gauge group $G$, $Z$ is the one-loop determinant. In \cite{KZ21}, the authors used the partition function $\mathcal{I}$ to get the effective superpotential $W^{3d}_{\text{eff}}(\sigma,m)$,
\begin{equation}\label{3}
    \mathcal{I}\sim \text{exp}\left(\frac{1}{\epsilon}W_{\text{eff}}(\sigma,m)\right)
\end{equation}
The one-loop determinant of the vector multiplet is
\begin{equation}\label{4}
Z_{\text{vec}}=\prod_{\alpha\in \Delta}e^{\frac{1}{8\beta_{2}}(\alpha\cdot \sigma)^{2}}(e^{i\alpha\cdot \sigma};q^{2})_{\infty}
\end{equation}
where the set of the roots of $G$ is denoted by $\Delta$. The one-loop determinant of the 
chiral multiplet with Neumann boundary condition is
\begin{equation}\label{5}
    Z_{\text{chi}}^{\text{Neu}}=\prod_{w\in \mathcal{R}}e^{\mathcal{E}(iw\cdot \sigma+r\beta_{2}+im)}(e^{-iw\cdot \sigma-im}q^{r};q^{2})_{\infty}^{-1},\qquad q=e^{-\beta_{2}}
\end{equation}
where the weight's set of the corresponding representation is denoted by $\mathcal{R}$, the $R$-charge of the scalar is in the chiral multiplet $r$, and
\begin{equation*}
    \mathcal{E}(x)=\dfrac{1}{8\beta_{2}}x^{2}-\dfrac{1}{4}x+\dfrac{\beta_{2}}{12}
\end{equation*}
$\beta_{1}$ is the fugacity of the rotation along $S^{1}$, $\beta_{2}$ is the $U(1)$ charge fugacity, $\beta l=(\beta_{1}+\beta_{2})l$ is the circumference of $S^{1}$. Finally, the 3d effective superpotential is given by without $\text{FI}$-term
\begin{equation}\label{6}
\begin{aligned}
W^{3d}_{\text{eff}}(\sigma,m)=&\dfrac{1}{\beta_{2}}\sum_{w \in \mathcal{R}}\sum_{a=1}^{N_{f}}\text{Li}_{2}(e^{-iw\cdot \sigma-im_{a}-i\beta_{2}\tilde{c}})-\dfrac{1}{4\beta_{2}}\sum_{w\in \mathcal{R}}\sum_{a=1}^{N_{f}}(w\cdot \sigma+m_{a}+\beta_{2}\tilde{c})^{2}\\
&-\dfrac{1}{\beta_{2}}\sum_{\alpha \in \Delta}\text{Li}_{2}(e^{i\alpha\cdot \sigma})+\dfrac{1}{4\beta_{2}}\sum_{\alpha\in \Delta}(\alpha\cdot \sigma)^{2}
\end{aligned}
\end{equation}
where $\beta_{2}$ is the $U(1)$ charge fugacity, the weight's set of the corresponding representation denoted by $\mathcal{R}$, the set of the roots of $G$ denoted by $\Delta$. The $\text{Li}_{2}(z)$ is called the dilogarithm,
\begin{equation*}
\text{Li}_{2}(z)=\sum_{k=1}^{\infty}\dfrac{z^{k}}{k^{2}}
\end{equation*}
We have known that the twisted effective superpotential $\widetilde{W}_{\text{eff}}(\sigma)$ contained two parts\cite{NS09a}: 
\begin{equation*}\label{}
    \widetilde{W}_{\text{eff}}(\sigma)=\widetilde{W}^{\text{matter}}_{\text{eff}}(\sigma)+\widetilde{W}^{\text{gauge}}_{\text{eff}}(\sigma)
\end{equation*}
The matter fields are generally in the chiral multiplets and the gauge fields are in the vector multiplet. So we can regard that the 3d $\mathcal{N}=2$ gauge theory contains matter field and gauge field. 

As we know, the vortex partition of 2d $\mathcal{N}=(2, 2)$ theory on the plane $\mathbb{C}$ with the $\Omega$ background can be obtained in the zero radius limit of the 3d partition function. And it can also alternatively
be computed as a dimensional reduction of 4d $\mathcal{N}=1$ gauge theory on the geometry.

\subsection{The new superpotential}

Following the calculation of effective superpotential in \cite{DZ23a}, we found a uniform 
method can also obtain this dual unexpectedly. In \cite{DZ23a}, we use the representation
\begin{equation*}
    \mathcal{R}=V\otimes V^{*}\oplus V\otimes \mathcal{F}\oplus V\otimes \tilde{\mathcal{F}}
\end{equation*}
to get the effective superpotential of $\text{BCD}$-type gauge theory. Let us recall the general effective superpotential
\begin{equation}\label{96}
    \begin{aligned}
    W^{3d}_{\text{eff}}(\sigma,m)=  &-\dfrac{1}{\beta_{2}}\sum_{\alpha \in \Delta}\frac{4}{\alpha_{i}^{2}}\text{Li}_{2}(e^{i\alpha\cdot \sigma})+\dfrac{1}{4\beta_{2}}\sum_{\alpha\in \Delta}\frac{4}{\alpha_{i}^{2}}(\alpha\cdot \sigma)^{2}\\
        &+\dfrac{1}{\beta_{2}}\sum_{w \in \mathcal{R}}\sum_{a=1}^{N_{f}}\text{Li}_{2}(e^{-iw\cdot \sigma-im_{a}-i\beta_{2}\tilde{c}})-\dfrac{1}{4\beta_{2}}\sum_{w\in \mathcal{R}}\sum_{a=1}^{N_{f}}(w\cdot \sigma+m_{a}+\beta_{2}\tilde{c})^{2}\\
    &+\dfrac{1}{\beta_{2}}\sum_{w \in \mathcal{R}}\sum_{a=1}^{N_{f}^{'}}\text{Li}_{2}(e^{iw\cdot \sigma-im_{a}^{'}-i\beta_{2}\tilde{c}})-\dfrac{1}{4\beta_{2}}\sum_{w\in \mathcal{R}}\sum_{a=1}^{N_{f}^{'}}(w\cdot \sigma-m_{a}^{'}-\beta_{2}\tilde{c})^{2}
   \end{aligned}
\end{equation}
where $\alpha_{i}$ is the root of Lie algebra of the Lie group. 

Here, if we choose another kind of representation 
\begin{equation}\label{90}
    \mathcal{R}=V\otimes V^{*}\oplus V\otimes \mathcal{F}
\end{equation}
where $V$ is the fundamental representation and $\mathcal{F}$ is the $N_{f}$ dimensional representation, and we obtain a new effective superpotential
\begin{equation}\label{7}
    \begin{aligned}
    W^{3d}_{\text{eff}}(\sigma,m)=&\dfrac{1}{\beta_{2}}\sum_{w \in \mathcal{R}}\sum_{a=1}^{N_{f}}\text{Li}_{2}(e^{2(-iw\cdot \sigma-im_{a}-i\beta_{2}\tilde{c})})-\dfrac{1}{4\beta_{2}}\sum_{w\in \mathcal{R}}\sum_{a=1}^{N_{f}}[2(w\cdot \sigma+m_{a}+\beta_{2}\tilde{c})]^{2}\\
    &-\dfrac{1}{\beta^{2}}\sum_{\alpha \in \Delta}\text{Li}_{2}(e^{\frac{4}{\alpha_{i}^{2}}i\alpha\cdot \sigma})+\dfrac{1}{4\beta_{2}}\sum_{\alpha\in \Delta}(\frac{4}{\alpha_{i}^{2}}\alpha\cdot \sigma)^{2}
   \end{aligned}
\end{equation}
Please note that we do not consider either the $\text{FI}$-term here. Comparing with the formula (\ref{6}), we can see that we put the weight factor $\dfrac{4}{\alpha_{i}^{2}}$ to the adjoint chiral multiplet, and twice the fundamental chiral multiplet. The weight factors in effective superpotential $W^{3d}_{\text{eff}}$ (\ref{7}) expands the range of application of (\ref{6}).

As an example, we calculate the $U(N)$ supersymmetry gauge theory. It is obvious that the root system consists of all $\{\sigma_{i}-\sigma_{j}\}$, $i\neq j$ \cite{Hum72}. We choose the representation
\begin{equation*}
    \mathcal{R}=V\otimes V^{*}\oplus V\otimes \mathcal{F}\oplus V\otimes \tilde{\mathcal{F}}
\end{equation*}
which corresponds to the theory with the $H^{\text{max}}=U(L)\times U(L)$ global symmetry group. Similarly, $V=\mathbf{C}^{N}$ is the $N$-dimensional fundamental representation, $\mathcal{F}\approx \mathbf{C}^{L}$, $\tilde{\mathcal{F}}\approx \mathbf{C}^{L}$ are the $L$-dimensional fundamental representations. In the following calculation, we absorb $\beta_{2}\tilde{c}$ into the mass parameters $m_{a}$. Then we get the new effective superpotential:
\begin{equation}\label{8}
\begin{aligned}
W^{A; 3d}_{\text{eff}}(\sigma,m)=&\dfrac{1}{\beta_{2}}\sum_{j\neq k}^{N}\text{Li}_{2}(e^{-2i(\sigma_{j}-\sigma_{k})-2im_{\text{adj}}})-\dfrac{1}{\beta_{2}}\sum_{j\neq k}^{N}(\sigma_{j}-\sigma_{k}+m_{\text{adj}})^{2}\\
&+\dfrac{1}{\beta_{2}}\sum_{j=1}^{N}\sum_{a=1}^{N_{f}}\text{Li}_{2}(e^{-2i \sigma_{j}-2im_{a}})-\dfrac{1}{\beta_{2}}\sum_{j=1}^{N}\sum_{a=1}^{N_{f}}( \sigma_{j}+m_{a})^{2}\\
&+\dfrac{1}{\beta_{2}}\sum_{j=1}^{N}\sum_{a=1}^{N_{f}^{'}}\text{Li}_{2}(e^{2i \sigma_{j}-2im_{a}^{'}})-\dfrac{1}{\beta_{2}}\sum_{j=1}^{N}\sum_{a=1}^{N_{f}^{'}}( \sigma_{j}-m_{a}^{'})^{2}\\
&-\dfrac{1}{\beta_{2}}\sum_{j\neq k}^{N}\text{Li}_{2}(e^{2i(\sigma_{j}-\sigma_{k})})+\dfrac{1}{\beta_{2}}\sum_{j\neq k}^{N}(\sigma_{j}-\sigma_{k})^{2}
\end{aligned}
\end{equation}
Using the formula (\ref{2}), the vacuum equation is given by
\begin{equation*}\label{}
\dfrac{\prod_{a=1}^{N_{f}}\text{sin}^{2}(\sigma_{j}-m_{a}^{'})}{\prod_{a=1}^{N_{f}^{'}}\text{sin}^{2}(\sigma_{j}+m_{a})}=\prod_{k\neq j}^{N}\dfrac{\text{sin}^{2}(\sigma_{j}-\sigma_{k}+m_{\text{adj}})}{\text{sin}^{2}(\sigma_{j}-\sigma_{k}-m_{\text{adj}})}
\end{equation*}
In the case $N_{f}=N_{f}^{'}$ and $m_{a}=m_{a}^{'}$, we can simplify it to
\begin{equation}\label{9}
\dfrac{\prod_{a=1}^{N_{f}}\text{sin}(\sigma_{j}-m_{a}^{'})}{\prod_{a=1}^{N_{f}^{'}}\text{sin}(\sigma_{j}+m_{a})}=\pm \prod_{k\neq j}^{N}\dfrac{\text{sin}(\sigma_{j}-\sigma_{k}+m_{\text{adj}})}{\text{sin}(\sigma_{j}-\sigma_{k}-m_{\text{adj}})}
\end{equation}
An important identity is given by
\begin{equation*}\label{}
    \text{exp}\left(\dfrac{\partial}{\partial x}\text{Li}_{2}(e^{\pm x})\right)=(1-e^{\pm x})^{\mp}
\end{equation*}

In \cite{DZ23a}, the representation of the $\text{A}$-type gauge theory is 
\begin{equation*}
    \mathcal{R}=V\otimes V^{*}\oplus V\otimes \mathcal{F}\oplus V\otimes \mathcal{F}\oplus V^{*}\otimes \tilde{\mathcal{F}}\oplus V^{*}\otimes \tilde{\mathcal{F}}
\end{equation*}
and obviously the representation in this article is chosen as
\begin{equation*}
    \mathcal{R}=V\otimes V^{*}\oplus V\otimes \mathcal{F}\oplus V\otimes \tilde{\mathcal{F}}
\end{equation*}
Comparing with the two representations and the effective superpotentials of $\text{SU}(N)$ supersymmetry gauge theory, the representation here only contains adjoint representation, and once fundamental representation, as well as once anti-fundamental. The reason is that we lift the weight factor to the index. With these variations, calculations become simpler. Nevertheless, that the two types of effective superpotential are different, while their vacuum equations are the same.

We know that the $\text{ADE}$-type Lie algebras are all simple-laced and self dual. But, please note that representation of the $\text{A}$-type gauge theory needs to contain the anti-fundamental part, which is different from the $\text{DE}$-type gauge theory. We will discuss the $\text{DE}$-type gauge theories in a later section.

\section{The superpotential of $\text{BCD}$-type Lie algebra}\label{b}

Please note that the new effective superpotential 
\begin{equation*}
    \begin{aligned}
    W^{3d}_{\text{eff}}(\sigma,m)=&\dfrac{1}{\beta_{2}}\sum_{w \in \mathcal{R}}\sum_{a=1}^{N_{f}}\text{Li}_{2}(e^{2(-iw\cdot \sigma-im_{a}-i\beta_{2}\tilde{c})})-\dfrac{1}{4\beta_{2}}\sum_{w\in \mathcal{R}}\sum_{a=1}^{N_{f}}[2(w\cdot \sigma+m_{a}+\beta_{2}\tilde{c})]^{2}\\
    &-\dfrac{1}{\beta^{2}}\sum_{\alpha \in \Delta}\text{Li}_{2}(e^{\frac{4}{\alpha_{i}^{2}}i\alpha\cdot \sigma})+\dfrac{1}{4\beta_{2}}\sum_{\alpha\in \Delta}(\frac{4}{\alpha_{i}^{2}}\alpha\cdot \sigma)^{2}
   \end{aligned}
\end{equation*}
The representation is 
\begin{equation*}\label{}
    \mathcal{R}=V\otimes V^{*}\oplus V\otimes \mathcal{F}
\end{equation*}

Following the same process, we can give the effective potential of $\text{BCD}$-type gauge theories. Then we use the formula (\ref{2}) to get the corresponding vacuum equations and discuss the connection between the gauge theory results and the open spin chains models based on this expression.

For the $\text{BCD}$-type gauge theory, we still choose the representation (\ref{99}). In the case of $\text{SO}(2N+1)$ gauge group, all the roots are given by $\{\pm \sigma_{i}\pm \sigma_{j}\}$ for all the possible combinations of $i<j$ and $\{\pm \sigma_{i}\}_{i=1}^{N}$. The effective superpotential is
\begin{equation}\label{10}
	\begin{aligned}
	W^{B_N, 3d}_{\text{eff}}(\sigma,m)=&-\dfrac{1}{\beta_{2}}\sum_{j<k}^{N}\text{Li}_{2}(e^{2i(\pm\sigma_{j}\pm \sigma_{k})})+\dfrac{1}{\beta_{2}}\sum_{j<k}^{N}(\pm \sigma_{j}\pm \sigma_{k})^{2}\\
	&+\dfrac{1}{\beta_{2}}\sum_{j<k}^{N}\text{Li}_{2}(e^{-2i(\pm \sigma_{j}\pm \sigma_{k})-2im_{\text{adj}}})-\dfrac{1}{\beta_{2}}\sum_{j<k}^{N}(\pm \sigma_{j}\pm \sigma_{k}+m_{\text{adj}})^{2}\\
	&+\dfrac{1}{\beta_{2}}\sum_{j=1}^{N}\sum_{a=1}^{N_{f}}\text{Li}_{2}(e^{-2(\pm i \sigma_{j}+im_{a})})-\dfrac{1}{\beta_{2}}\sum_{j=1}^{N}\sum_{a=1}^{N_{f}}(\pm \sigma_{j}+m_{a})^{2}\\
&-\dfrac{1}{\beta_{2}}\sum_{j=1}^{N}\text{Li}_{2}(e^{\pm 4i \sigma_{j}})+\dfrac{8}{\beta_{2}}\sum_{j=1}^{N}\sigma_{j}^{2}\\
&+\dfrac{1}{\beta_{2}}\sum_{j=1}^{N}\text{Li}_{2}(e^{(\pm 4i \sigma_{j}-4im_{\text{adj}})})-\dfrac{4}{\beta_{2}}\sum_{j=1}^{N}(\sigma_{j}\pm m_{\text{adj}})^{2}
\end{aligned}
\end{equation}
Using the vacuum equation (\ref{2}), the vacuum equation is read
\begin{equation}\label{11}
\begin{aligned}
    &\dfrac{\text{sin}^{4}(2\sigma_{j}-2m_{\text{adj}})}{\text{sin}^{4}(2\sigma_{j}+2m_{\text{adj}})}\prod_{j\neq k}\dfrac{\text{sin}^{2}(\sigma_{j}\pm \sigma_{k}-m_{\text{adj}})}{\text{sin}^{2}(-\sigma_{j}\pm \sigma_{k}-m_{\text{adj}})} \prod_{a=1}^{N_{f}}\dfrac{\text{sin}^{2}(\sigma_{j}-m_{a})}{\text{sin}^{2}(-\sigma_{j}-m_{a})}=1
\end{aligned}
\end{equation}
Note that 
\begin{equation*}
    \dfrac{\text{sin}^{2}(2\sigma_{j}-2m_{\text{adj}})}{\text{sin}^{2}(2\sigma_{j}+m_{\text{adj}})}=\dfrac{\text{sin}^{2}(\sigma_{j}-m_{\text{adj}})\text{cos}^{2}(\sigma_{j}-m_{\text{adj}})}{\text{sin}^{2}(\sigma_{j}+m_{\text{adj}})\text{cos}^{2}(\sigma_{j}+m_{\text{adj}})}
\end{equation*}
Then we get two type vacuum equations by taking the square root of (\ref{11})
\begin{equation}\label{12}
    \dfrac{\text{sin}^{2}(\sigma_{j}-m_{\text{adj}})\text{cos}^{2}(\sigma_{j}-m_{\text{adj}})}{\text{sin}^{2}(\sigma_{j}+m_{\text{adj}})\text{cos}^{2}(\sigma_{j}+m_{\text{adj}})}\prod_{j\neq k}\dfrac{\text{sin}(\sigma_{j}\pm \sigma_{k}-m_{\text{adj}})}{\text{sin}(-\sigma_{j}\pm \sigma_{k}-m_{\text{adj}})}\prod_{a=1}^{N_{f}}\dfrac{\text{sin}(\sigma_{j}-m_{a})}{\text{sin}(-\sigma_{j}-m_{a})}=\pm 1
\end{equation}

For $\text{Sp}(2N)$ gauge theory, all the roots are given by $\{\pm \sigma_{i}\pm \sigma_{j}\}$ for $i<j$ and $\{\pm 2\sigma_{i}\}_{i=1}^{N}$. The effective superpotential is
\begin{equation}\label{14}
	\begin{aligned}
	W^{C_N, 3d}_{\text{eff}}(\sigma,m)=&-\dfrac{1}{\beta_{2}}\sum_{j<k}^{N}\text{Li}_{2}(e^{2i(\pm\sigma_{j}\pm \sigma_{k})})+\dfrac{1}{\beta_{2}}\sum_{j<k}^{N}(\pm \sigma_{j}\pm \sigma_{k})^{2}\\
&+\dfrac{1}{\beta_{2}}\sum_{j<k}^{N}\text{Li}_{2}(e^{-2i(\pm \sigma_{j}\pm \sigma_{k})-2im_{\text{adj}}})-\dfrac{1}{\beta_{2}}\sum_{j<k}^{N}(\pm \sigma_{j}\pm \sigma_{k}+m_{\text{adj}})^{2}\\
&+\dfrac{1}{\beta_{2}}\sum_{j=1}^{N}\sum_{a=1}^{N_{f}}\text{Li}_{2}(e^{-2(\pm i \sigma_{j}+im_{a})})-\dfrac{1}{\beta_{2}}\sum_{j=1}^{N}\sum_{a=1}^{N_{f}}(\pm \sigma_{j}+m_{a})^{2}\\
&-\dfrac{1}{\beta_{2}}\sum_{j=1}^{N}\text{Li}_{2}(e^{\pm 2i \sigma_{j}})+\dfrac{2}{\beta_{2}}\sum_{j=1}^{N}\sigma_{j}^{2}
\end{aligned}
\end{equation}
The vacuum equation is given by (\ref{2})
\begin{equation}\label{15}
\begin{aligned}
    &\dfrac{\text{sin}^{2}(\sigma_{j}-\frac{m_{\text{adj}}}{2})}{\text{sin}^{2}(\sigma_{j}+\frac{m_{\text{adj}}}{2})}\prod_{j\neq k}\dfrac{\text{sin}^{2}(\sigma_{j}\pm \sigma_{k}-m_{\text{adj}})}{\text{sin}^{2}(-\sigma_{j}\pm \sigma_{k}-m_{\text{adj}})}\prod_{a=1}^{N_{f}}\dfrac{\text{sin}^{2}(\sigma_{j}-m_{a})}{\text{sin}^{2}(-\sigma_{j}-m_{a})}=1
\end{aligned}
\end{equation}
Then we get two vacuum equations by square root of (\ref{15})
\begin{equation}\label{16}
    \dfrac{\text{sin}(\sigma_{j}-\frac{m_{\text{adj}}}{2})}{\text{sin}(\sigma_{j}+\frac{m_{\text{adj}}}{2})}\prod_{j\neq k}\dfrac{\text{sin}(\sigma_{j}\pm \sigma_{k}-m_{\text{adj}})}{\text{sin}(-\sigma_{j}\pm \sigma_{k}-m_{\text{adj}})}\prod_{a=1}^{N_{f}}\dfrac{\text{sin}(\sigma_{j}-m_{a})}{\text{sin}(-\sigma_{j}-m_{a})}=\pm 1
\end{equation}

In the case of $\text{SO}(2N)$ gauge group, all the roots are given by $\{\pm \sigma_{i}\pm \sigma_{j}\}$ for all $1\le i< j\le N$. The effective superpotential is  
\begin{equation}\label{18}
\begin{aligned}
W^{D_N, 3d}_{\text{eff}}(\sigma,m)=&-\dfrac{}{\beta_{2}}\sum_{j<k}^{N}\text{Li}_{2}(e^{2i(\pm\sigma_{j}\pm \sigma_{k})})+\dfrac{1}{\beta_{2}}\sum_{j<k}^{N}(\pm \sigma_{j}\pm \sigma_{k})^{2}\\
&+\dfrac{1}{\beta_{2}}\sum_{j<k}^{N}\text{Li}_{2}(e^{-2i(\pm \sigma_{j}\pm \sigma_{k})-2im_{\text{adj}}})-\dfrac{1}{\beta_{2}}\sum_{j<k}^{N}(\pm \sigma_{j}\pm \sigma_{k}+m_{\text{adj}})^{2}\\
&+\dfrac{1}{\beta_{2}}\sum_{j=1}^{N}\sum_{a=1}^{N_{f}}\text{Li}_{2}(e^{-2(\pm i \sigma_{j}+im_{a})})-\dfrac{1}{\beta_{2}}\sum_{j=1}^{N}\sum_{a=1}^{N_{f}}(\pm \sigma_{j}+m_{a})^{2}
\end{aligned}
\end{equation}
The vacuum equation is 
\begin{equation}\label{19}
\begin{aligned}
    \prod_{j\neq k}\dfrac{\text{sin}^{2}(\sigma_{j}\pm \sigma_{k}-m_{\text{adj}})}{\text{sin}^{2}(-\sigma_{j}\pm \sigma_{k}-m_{\text{adj}})} \prod_{a=1}^{N_{f}}\dfrac{\text{sin}^{2}(\sigma_{j}-m_{a})}{\text{sin}^{2}(-\sigma_{j}-m_{a})}=1
\end{aligned}
\end{equation}
Then we get two type vacuum equations with square root of (\ref{19})
\begin{equation}\label{20}
    \prod_{j\neq k}\dfrac{\text{sin}(\sigma_{j}\pm \sigma_{k}-m_{\text{adj}})}{\text{sin}(-\sigma_{j}\pm \sigma_{k}-m_{\text{adj}})} \prod_{a=1}^{N_{f}}\dfrac{\text{sin}(\sigma_{j}-m_{a})}{\text{sin}(-\sigma_{j}-m_{a})}=\pm 1
\end{equation}
For all of these representations, the two branches of the vacuum equations are not connected. Without specifically mentioned, we use the positive one.

One can alternatively take a 2d limit of the 3d effective potential (\ref{7}) to derive the 
effective potential of 2d $\mathcal{N}=(2, 2)$ theories. In 2d limits, the $B$-, $C$- and 
$D$-type vacuum equations are degenerated to 
\begin{equation}\label{21}
    \dfrac{(\sigma_{j}-m_{\text{adj}})^{2}}{(\sigma_{j}+m_{\text{adj}})^{2}}\prod_{j\neq k}\dfrac{\sigma_{j}\pm \sigma_{k}-m_{\text{adj}}}{-\sigma_{j}\pm \sigma_{k}-m_{\text{adj}}}\prod_{a=1}^{N_{f}}\dfrac{\sigma_{j}-m_{a}}{-\sigma_{j}-m_{a}}=\pm1~~~(B)
\end{equation}
\begin{equation}\label{22}
    \dfrac{\sigma_{j}-\frac{m_{\text{adj}}}{2}}{\sigma_{j}+\frac{m_{\text{adj}}}{2}}\prod_{j\neq k}\dfrac{\sigma_{j}\pm \sigma_{k}-m_{\text{adj}}}{-\sigma_{j}\pm \sigma_{k}-m_{\text{adj}}}\prod_{a=1}^{N_{f}}\dfrac{\sigma_{j}-m_{a}}{-\sigma_{j}-m_{a}}=\pm 1~~~(C)
\end{equation}
and 
\begin{equation}\label{23}
    \prod_{j\neq k}\dfrac{\sigma_{j}\pm \sigma_{k}-m_{\text{adj}}}{-\sigma_{j}\pm \sigma_{k}-m_{\text{adj}}} \prod_{a=1}^{N_{f}}\dfrac{\sigma_{j}-m_{a}}{-\sigma_{j}-m_{a}}=\pm 1~~~(D),
\end{equation}
respectively.

\section{Bethe/Gauge correspondence}\label{c}

In this section, we present the explicit atlas between the vacuum equation and the Bethe Ansatz equation, relating the 3d $\mathcal{N}=2$ supersymmetric gauge theories on $S^{1}\times D^{2}$ and the quantum integrable spin chains. 

The Bethe ansatz equation with periodic boundary condition is
\begin{equation}\label{24}
\prod_{a=1}^{L}\dfrac{[u_{i}+\frac{\eta}{2}+\eta s_{a}-\vartheta_{a}]}{[u_{i}+\frac{\eta}{2}-\eta s_{a}-\vartheta_{a}]}=\prod_{j\neq i,j=1}^{M}\dfrac{[u_{i}-u_{j}+\eta]}{[u_{i}-u_{j}-\eta]}
\end{equation}
The Bethe ansatz equation of the $\text{XXZ}$ spin chain with diagonal boundary condition is given by
\begin{equation}\label{25}
\begin{aligned}
&\dfrac{\text{sin}[\pi (u_{i}-\frac{\eta}{2}+\xi_{+})]}{\text{sin}[\pi(u_{i}+\frac{\eta}{2}-\xi_{+})]}\dfrac{\text{sin}[\pi (u_{i}-\frac{\eta}{2}+\xi_{-})]}{\text{sin}[\pi(u_{i}+\frac{\eta}{2}-\xi_{-})]}\\
&\times\prod_{a=1}^{L}\dfrac{\text{sin}[\pi (u_{i}+\frac{\eta}{2}+\eta s_{a}-\vartheta_{a})]\text{sin}[\pi(-u_{i}+\frac{\eta}{2}-\eta s_{a}-\vartheta_{a})]}{\text{sin}[\pi(-u_{i}+\frac{\eta}{2}+\eta s_{a}-\vartheta_{a})]\text{sin}[\pi(u_{i}+\frac{\eta}{2}-\eta s_{a}-\vartheta_{a})]}\\
&\times \prod_{j\neq i,j=1}^{M}\dfrac{\text{sin}[\pi (u_{j}+u_{i}-\eta)]\text{sin}[\pi(u_{j}-u_{i}-\eta])}{\text{sin}[\pi(u_{j}-u_{i}+\eta)]\text{sin}[\pi(u_{j}+u_{i}+\eta)]}=1
\end{aligned}
\end{equation}
In the 2d limit, the $\text{XXZ}$ spin chain is changed to $\text{XXX}$ spin chain, where $[u]\rightarrow u$. The degenerated Bethe ansatz equation of the $\text{XXX}$ spin chain with diagonal boundary condition is 
\begin{equation}\label{26}
\begin{aligned}
&\dfrac{(u_{j}-\frac{\eta}{2}+\xi_{+})}{(u_{j}+\frac{\eta}{2}-\xi_{+})}\dfrac{(u_{j}-\frac{\eta}{2}+\xi_{-})}{(u_{j}+\frac{\eta}{2}-\xi_{-})}\prod_{a=1}^{L}\dfrac{(u_{j}+\frac{\eta}{2}+\eta s_{a}-\vartheta_{a})(-u_{j}+\frac{\eta}{2}-\eta s_{a}-\vartheta_{a})}{(-u_{j}+\frac{\eta}{2}+\eta s_{a}-\vartheta_{a})(u_{j}+\frac{\eta}{2}-\eta s_{a}-\vartheta_{a})}\\
&\times \prod_{k\neq j,k=1}^{M}\dfrac{ (u_{k}-u_{j}-\eta)(u_{k}+u_{j}-\eta)}{(u_{k}-u_{j}+\eta)(u_{k}+u_{j}+\eta)}=1
\end{aligned}
\end{equation} 
In appendix \ref{A}, we list a more detailed review on how to derive the above Bethe ansatz equations.

\subsection{$A_{N}$-type gauge theory}

For $A$-type gauge theory, we can see that the vacuum equation (\ref{9}) is matched with the Bethe equation (\ref{24}). The dictionary is given by
\begin{equation}\label{}
\begin{aligned}
&\pi u \longleftrightarrow \sigma ,\quad \pi \eta \longleftrightarrow m_{\text{adj}}\\
&M \longleftrightarrow N,\quad L\longleftrightarrow N_{f}=N_{f}^{'}\\
&\{-\pi \eta s_{a}-\dfrac{\pi \eta}{2} +\pi \vartheta_{a}\}\longleftrightarrow m_{a}^{'},\quad \{-\pi \eta s_{a}+\dfrac{\pi \eta}{2} -\pi \vartheta_{a}\} \longleftrightarrow  m_{a}
\end{aligned}
\end{equation}
which is the same to the dictionary of $\text{A}$-type gauge theory in \cite{DZ23a}. 

\subsection{$B_{N}$-type gauge theory}

For $B_{N}$ gauge group, we can compare the positive branch of the vacuum equation (\ref{12}) and the Bethe equation (\ref{25}). Then we find that there is more than one solution which can satisfy the duality. A choice of the boundary parameters can be 
$$\xi_{+}=\xi_{-}=-\dfrac{\eta}{2}+\dfrac{1}{2}$$
the spins should be 
$s_{1}=s_{2}=-\dfrac{1}{2}$, 
and the inhomogeneous parameters are $\vartheta_{1}=\vartheta_{2}=0$. Next, the boundary parameters can be
$$\xi_{+}=\xi_{-}=-\dfrac{\eta}{2}$$
the spins should be $s_{1}=s_{2}=-\dfrac{1}{2}$, and $\vartheta_{1}=\vartheta_{2}=\dfrac{1}{2}$. On both of those occasions, we see the fact that $N_{f}=2(L-2)$ is an even integer and the map is 
\begin{equation}\label{99}
\begin{aligned}
&\pi u \longleftrightarrow \sigma ,\quad \pi \eta \longleftrightarrow m_{\text{adj}}\\
& M \longleftrightarrow N,\quad 2(L-2)\longleftrightarrow N_{f} \\
&\{-\pi \eta s_{a}-\dfrac{\pi \eta}{2} +\pi \vartheta_{a},-\pi \eta s_{a}+\dfrac{\pi \eta}{2} -\pi \vartheta_{a}\} \longleftrightarrow  m_{a}
 \end{aligned}
\end{equation}
Another choice of the boundary parameters can be 
$$\xi_{+}=\xi_{-}=\dfrac{\eta}{2}$$
then, the spins should be $s_{1}=s_{2}=s_{3}=s_{4}=-\dfrac{1}{2}$, the $\vartheta_{1}=\vartheta_{2}=0$, and $\vartheta_{3}=\vartheta_{4}=\dfrac{1}{2}$. In this case, the $N_{f}=2(L-4)$, and other dictionaries are the same to (\ref{99}). The third choice of the boundary parameters can be
$$\xi_{+}=-\dfrac{\eta}{2}, \qquad  \xi_{-}=\dfrac{\eta}{2}$$
the spins should be $s_{1}=s_{2}=s_{3}=-\dfrac{1}{2}$, the $\vartheta_{1}=0$ and $\vartheta_{2}=\vartheta_{3}=\dfrac{1}{2}$. 

The most interesting case is that, if we choose 
$$\xi_{+}=-\dfrac{\eta}{2}+\dfrac{1}{2}, \qquad  \xi_{-}=\dfrac{\eta}{2}$$
the spins $s_{1}=s_{2}=s_{3}=-\dfrac{1}{2}$, and the $\vartheta_{1}=\dfrac{1}{2}$ and $\vartheta_{2}=\vartheta_{3}=0$. The positive branch of the vacuum equations will change to the negative one, which corresponds to the same Bethe equations. In these two cases, we have $N_{f}=2(L-3)$, and other parameters are the same to (\ref{99}).

Except for the disappearance $N_{f}^{'}$, the map is also the same to the dictionary as the $\text{B}$-type gauge theory in \cite{DZ23a}. The spin $s=-\dfrac{1}{2}$ is the most difference from the results of $\text{B}$-type gauge theory in \cite{DZ23a}. We will explain this point in detail in section \ref{e}. Actually, as the vacuum equations and the Bethe equations are both trigonometric function, so the above parameters are not unique.

\subsection{$C_{N}$-type gauge theory}

For $C_{N}$ gauge group, we can compare the vacuum equation (\ref{16}) and the Bethe equation (\ref{25}). We choose one of the following boundary conditions
\begin{itemize}
\item \qquad $\xi_{+}=\dfrac{\eta}{2},\qquad \xi_{-}=0$
\item \qquad $\xi_{+}=0,\qquad \xi_{-}=\dfrac{\eta}{2}$
\end{itemize}
to get the correspondence. The dictionary is the same to (\ref{99}). And the boundary parameters above are different from the  values in \cite{DZ23a}.

\subsection{$D_{N}$-type gauge theory}

For $D_{N}$ gauge group, we can compare the vacuum equation (\ref{20}) and the Bethe equation (\ref{25}). The same dictionary (\ref{99}) maps the vacuum equation to the Bethe ansatz equation with the boundary parameters $\xi_{+}=\xi_{-}=i\infty$. It is the same to the results of the $\text{D}$-type gauge theory in \cite{DZ23a}.

\subsection{2d limits}

In 2d limits, the vacuum equations and Bethe equations are replaced by the rational analogues. For $B_{N}$-type supersymmetry gauge theory, the vacuum equation (\ref{11}) degenerates to
\begin{equation}\label{27}
\begin{aligned}
    &\dfrac{(\sigma_{j}-m_{\text{adj}})^{4}}{(\sigma_{j}+m_{\text{adj}})^{4}}\prod_{j\neq k}\dfrac{(\sigma_{j}\pm \sigma_{k}-m_{\text{adj}})^{2}}{(-\sigma_{j}\pm \sigma_{k}-m_{\text{adj}})^{2}} \prod_{a=1}^{N_{f}}\dfrac{(\sigma_{j}-m_{a})^{2}}{(-\sigma_{j}-m_{a})^{2}}=1
\end{aligned}
\end{equation}
So we can choose the boundary parameters $\xi_{+}=\xi_{-}=-\dfrac{\eta}{2}$ to match the positive branch of the above vacuum equation (\ref{21}). The precise relation reads as follows:
\begin{equation}\label{98}
\begin{aligned}
&u \longleftrightarrow \sigma ,\quad \eta \longleftrightarrow m_{\text{adj}}\\
& M \longleftrightarrow N,\quad 2L\longleftrightarrow N_{f}\\
&\{-\eta s_{a}-\dfrac{\eta}{2} +\vartheta_{a},-\eta s_{a}+\dfrac{ \eta}{2} -\vartheta_{a}\} \longleftrightarrow  m_{a}
 \end{aligned}
\end{equation}

For $C_{N}$-type supersymmetry gauge theory, the vacuum equation (\ref{15}) degenerates to
\begin{equation}\label{28}
\begin{aligned}
    &\dfrac{(\sigma_{j}-\frac{m_{\text{adj}}}{2})^2}{(\sigma_{j}+\frac{m_{\text{adj}}}{2})^2}\prod_{j\neq k}\dfrac{(\sigma_{j}\pm \sigma_{k}-m_{\text{adj}})^2}{(-\sigma_{j}\pm \sigma_{k}-m_{\text{adj}})^2}\prod_{a=1}^{N_{f}}\dfrac{(\sigma_{j}-m_{a})^2}{(-\sigma_{j}-m_{a})^2}=1
\end{aligned}
\end{equation}
With the same dictionary (\ref{98}), we choose one of the following two kinds of boundary conditions to match the positive part of the vacuum equation (\ref{22}).
\begin{itemize}
\item \qquad $\xi_{+}=\dfrac{\eta}{2},\qquad \xi_{-}=0$
\item \qquad $\xi_{+}=0,\qquad \xi_{-}=\dfrac{\eta}{2}$
\end{itemize}

For $D_{N}$-type supersymmetry gauge theory, the vacuum equation (\ref{19}) degenerates to
\begin{equation}\label{29}
\begin{aligned}
    \prod_{j\neq k}\dfrac{(\sigma_{j}\pm \sigma_{k}-m_{\text{adj}})^2}{(-\sigma_{j}\pm \sigma_{k}-m_{\text{adj}})^2} \prod_{a=1}^{N_{f}}\dfrac{(\sigma_{j}-m_{a})^2}{(-\sigma_{j}-m_{a})^2}=1
\end{aligned}
\end{equation}
The boundary condition should be taken $\xi_{+}=\xi_{-}=\dfrac{\eta}{2}$ to fit the positive term of the vacuum equation (\ref{23}). The dictionary is the same to (\ref{98}).

\section{Exception Lie algebras}\label{d}

Except for classical Lie algebras $\text{A}_{N}$, $\text{B}_{N}$, $\text{C}_{N}$, $\text{D}_{N}$, we also consider exceptional Lie algebras $\text{E}_{6},\text{E}_{7},\text{E}_{8},\text{F}_{4}$. The process for calculating the effective superpotential $W_{\text{eff}}^{3d}(\sigma,m)$ is the same. The root system of $\text{G}_{2}$ consists of \cite{Hum72}
$$\pm\left\{\sigma_{1}-\sigma_{2},\sigma_{2}-\sigma_{3},\sigma_{1}-\sigma_{3},2\sigma_{1}-\sigma_{2}-\sigma_{3},2\sigma_{2}-\sigma_{1}-\sigma_{3},2\sigma_{3}-\sigma_{1}-\sigma_{2}\right\}$$ Then we get $\dfrac{4}{\alpha_{i}}$ are 2, $\dfrac{2}{3}$. The appearance of fractional will make the moduli space for this case more complex, for convenience, we leave the discussion for the case of $\text{G}_{2}$ in the future.

\subsection{$\text{E}_{8}$-type}

We know that $\text{E}_{6}$, $\text{E}_{7}$ can be identified canonically with subsystems of $\text{E}_{8}$, so it is suffices to construct $E_{8}$. The root system of $\text{E}_{8}$ consists of the obvious vectors $\pm(e_{i}\pm e_{j})$, $i\neq j$, along with the less obvious ones $\frac{1}{2}\sum_{i=1}^{8}(-1)^{\kappa(i)}e_{i}$ (where the $\kappa(i)=0,1$, add up to an even integer) \cite{Hum72}. By using the result (\ref{8}), we calculate the effective superpotential
\begin{equation}\label{30}
\begin{aligned}
W_{\text{eff}}^{E_8, 3d}(\sigma,m)=&-\dfrac{1}{\beta_{2}}\sum_{j\neq k}^{8}\text{Li}_{2}(e^{2i(\pm\sigma_{j}\pm\sigma_{k})})+\dfrac{1}{\beta_{2}}\sum_{j\neq k}^{8}(\pm\sigma_{j}\pm \sigma_{k})^{2}\\
&+\dfrac{1}{\beta_{2}}\sum_{j\neq k}^{8}\text{Li}_{2}(e^{-2i(\pm\sigma_{j}\pm\sigma_{k})-im_{\text{adj}}})-\dfrac{1}{\beta_{2}}\sum_{j\neq k}^{8}(\pm\sigma_{j}\pm \sigma_{k}+m_{\text{adj}})^{2}\\
&-\dfrac{1}{\beta_{2}}\text{Li}_{2}(e^{2i(\frac{1}{2}\sum_{j=1}^{8}(-1)^{\kappa(j)}\sigma_{j})})+\dfrac{1}{\beta_{2}}(\sum_{j=1}^{8}(-1)^{\kappa(j)}\sigma_{j})^{2}\\
&+\dfrac{1}{\beta_{2}}\text{Li}_{2}(e^{-2i(\frac{1}{2}\sum_{j=1}^{8}(-1)^{\kappa(j)}\sigma_{j})-im_{\text{adj}}})-\dfrac{1}{\beta_{2}}(\sum_{j=1}^{8}(-1)^{\kappa(j)}\sigma_{j}+m_{\text{adj}})^{2}\\
&+\dfrac{1}{\beta_{2}}\sum_{j=1}^{8}\sum_{a=1}^{N_{f}}\text{Li}_{2}(e^{-2i(\pm \sigma_{j}+m_{a})})-\dfrac{1}{\beta_{2}}\sum_{j=1}^{8}\sum_{a=1}^{N_{f}}(\pm \sigma_{j}+m_{a})^{2}
\end{aligned}
\end{equation}
It is easy to know that the root system $\text{E}_{8}$ has 240 roots. So there are 240 vacuum equations in $\text{E}_{8}$-type totally. One can use the formula (\ref{2}) to get the explicit vacuum equations. The root system $\text{E}_{7}$ has 126 roots and the root system $\text{E}_{6}$ has 72 roots. We can see that the effective superpotentials (\ref{30}) here are different from the formula of the $\text{E}_{8}$-type in \cite{DZ23a}.

\subsection{$\text{F}_{4}$-type}

The root system of $\text{F}_{4}$ consists of all $\pm e_{i}$, all $\pm(e_{i}\pm e_{j})$, $i\neq j$, as well as all $\pm \frac{1}{2}(e_{1}\pm e_{2}\pm e_{3}\pm e_{4})$, where the signs may be chosen independently. Using the formula (\ref{8}), we get the effective superpotential 
\begin{equation}\label{31}
\begin{aligned}
W_{\text{eff}}^{F_4, 3d}(\sigma,m)=&-\dfrac{1}{\beta_{2}}\sum_{j\neq k}^{4}\text{Li}_{2}(e^{2i(\pm\sigma_{j}\pm\sigma_{k})})+\dfrac{1}{\beta_{2}}\sum_{j\neq k}^{4}(\pm\sigma_{j}\pm \sigma_{k})^{2}\\
&+\dfrac{1}{\beta_{2}}\sum_{j\neq k}^{4}\text{Li}_{2}(e^{-2i(\pm\sigma_{j}\pm\sigma_{k})-im_{\text{adj}}})-\dfrac{1}{\beta_{2}}\sum_{j\neq k}^{4}(\pm\sigma_{j}\pm \sigma_{k}+m_{\text{adj}})^{2}\\
&-\dfrac{1}{\beta_{2}}\sum_{j=1}^{4}\text{Li}_{2}(e^{\pm 4i \sigma_{j}})+\dfrac{4}{\beta_{2}}\sum_{j=1}^{4}(\sigma_{j})^{2}\\
&+\dfrac{1}{\beta_{2}}\sum_{j=1}^{4}\text{Li}_{2}(e^{\pm 4i\sigma_{j}-im_{\text{adj}}})-\dfrac{4}{\beta_{2}}\sum_{j=1}^{4}(\sigma_{j}\pm m_{\text{adj}})^{2}\\
&-\dfrac{1}{\beta_{2}}\text{Li}_{2}(e^{4i(\pm \frac{1}{2}(\sigma_{1}\pm \sigma_{2}\pm \sigma_{3}\pm \sigma_{4}))})+\dfrac{1}{\beta_{2}}(\sigma_{1}\pm \sigma_{2}\pm \sigma_{3}\pm\sigma_{4})^{2}\\
&+\dfrac{1}{\beta_{2}}\text{Li}_{2}(e^{-4i(\pm \frac{1}{2}(\sigma_{1}\pm \sigma_{2}\pm \sigma_{3}\pm \sigma_{4}))-im_{\text{adj}}})-\dfrac{1}{\beta_{2}}(\sigma_{1}\pm \sigma_{2}\pm \sigma_{3}\pm\sigma_{4}\pm m_{\text{adj}})^{2}\\
&+\dfrac{1}{\beta_{2}}\sum_{j=1}^{4}\sum_{a=1}^{N_{f}}\text{Li}_{2}(e^{-4i(\pm \sigma_{j}+m_{a})})-\dfrac{1}{\beta_{2}}\sum_{j=1}^{4}\sum_{a=1}^{N_{f}}(\pm \sigma_{j}+m_{a})^{2}
\end{aligned}
\end{equation}
The vacuum equations are given by using the formula (\ref{2})
\begin{equation}\label{32}
\begin{aligned}
&\dfrac{\text{sin}^{4}(\sigma_{j}-m_{\text{adj}})}{\text{sin}^{4}(\sigma_{j}+m_{\text{adj}})}\prod_{k\neq j}^{4}\dfrac{\text{sin}^{2}(\sigma_{j}\pm \sigma_{k}-m_{\text{adj}})}{\text{sin}^{2}(-\sigma_{j}\pm \sigma_{k}-m_{\text{adj}})}\\
&\times \dfrac{\text{sin}^{2}(\sigma_{j}\pm \sum_{k\neq j}^{4}\sigma_{k}-m_{\text{adj}})}{\text{sin}^{2}(-\sigma_{j}\pm \sum_{k\neq j}^{4}\sigma_{k}-m_{\text{adj}})}\dfrac{\text{sin}^{2}(\sigma_{1}-m_{a})}{\text{sin}^{2}(-\sigma_{1}-m_{a})}=1
\end{aligned}
\end{equation}
where $j,k=1,2,3,4$. So for $\text{F}_{4}$ Lie algebra, there are 4 vacuum equations. We can write the vacuum equation with the square root of (\ref{32})
\begin{equation*}\label{}
\begin{aligned}
&\dfrac{\text{sin}^{2}(\sigma_{j}-m_{\text{adj}})}{\text{sin}^{2}(\sigma_{j}+m_{\text{adj}})}\prod_{k\neq j}^{4}\dfrac{\text{sin}(\sigma_{j}\pm \sigma_{k}-m_{\text{adj}})}{\text{sin}(-\sigma_{j}\pm \sigma_{k}-m_{\text{adj}})}\\
&\times \dfrac{\text{sin}(\sigma_{j}\pm \sum_{k\neq j}^{4}\sigma_{k}-m_{\text{adj}})}{\text{sin}(-\sigma_{j}\pm \sum_{k\neq j}^{4}\sigma_{k}-m_{\text{adj}})}\dfrac{\text{sin}(\sigma_{1}-m_{a})}{\text{sin}(-\sigma_{1}-m_{a})}=\pm 1
\end{aligned}
\end{equation*}
It is also easy to notice that the effective superpotential and vacuum equation (\ref{32}) are different from the results in \cite{DZ23a}. As we let $N_{f}=N_{f}^{'}$, the square root of the vacuum equation in \cite{DZ23a} are the same as the square root of the vacuum equations (\ref{32}).

\section{Duality in the two effective superpotentials}\label{e}

In \cite{DZ23a}, we give a unified correspondence between 2d or 3d $\text{A}$-type supersymmetry gauge theory and closed $\text{XXX}$ spin chain, and a uniform correspondence between the $\text{BCD}$-type supersymmetry gauge theory and the $\text{XXZ}$ open spin chain with diagonal boundary condition. Given a $\text{ABCD}$-type gauge theory, the corresponding boundary conditions of the $\text{XXX}$ spin chain are different to each other, and  the vacuum equations can degenerate to the previous results at appropriate values. 

In \cite{DZ23a}, we chose the representation for $\text{BCD}$-type gauge theory
\begin{equation}\label{}
    \mathcal{R}^{'}=V\otimes V^{*}\oplus V\otimes \mathcal{F}\oplus V\otimes \mathcal{F}^{'}
\end{equation}
For $\text{B}$-type Lie group, we choose the vacuum equations
\begin{equation*}
\dfrac{\text{sin}^{2}(\sigma_{j}-m_{\text{adj}})}{\text{sin}^{2}(\sigma_{j}+m_{\text{adj}})}\prod_{j\neq k}^{N}\dfrac{\text{sin}(\sigma_{j}\pm \sigma_{k}-m_{\text{adj}})}{\text{sin}(-\sigma_{j}\pm \sigma_{k}-m_{\text{adj}})}\prod_{a=1}^{N_{f}}\dfrac{\text{sin}(\sigma_{j}-m_{a})}{\text{sin}(-\sigma_{j}-m_{a})}=1
\end{equation*}
to match the Bethe equation (\ref{25}). For $\text{C}$-type Lie group, on the other hand, the vacuum equation
\begin{equation*}
\dfrac{\text{sin}(\sigma_{j}-\frac{m_{\text{adj}}}{2})\text{cos}(\sigma_{j}-\frac{m_{\text{adj}}}{2})}{\text{sin}(\sigma_{j}+\frac{m_{\text{adj}}}{2})\text{cos}(\sigma_{j}+\frac{m_{\text{adj}}}{2})}\prod_{j\neq k}^{N}\dfrac{\text{sin}(\sigma_{j}\pm \sigma_{k}-m_{\text{adj}})}{\text{sin}(-\sigma_{j}\pm \sigma_{k}-m_{\text{adj}})}\prod_{a=1}^{N_{f}}\dfrac{\text{sin}(\sigma_{j}-m_{a})}{\text{sin}(-\sigma_{j}-m_{a})}=-1
\end{equation*}
were being used to fit the Bethe equation (\ref{25}). That is to say, we have to choose a different branch vacuum equation to get Bethe/Gauge correspondence for the $\text{B}$-type and $\text{C}$-type gauge theory. It is well known that, the $\text{B}$-type Lie groups Langlands dual to the $\text{C}$-type Lie groups. From this point, we can regard the $\text{B}$-type gauge theory and the $\text{C}$-type gauge theory duality to each other. For $\text{D}$-type gauge theory, the vacuum equations 
\begin{equation*}\label{}
\prod_{j\neq k}^{N}\dfrac{\text{sin}(\sigma_{j}\pm \sigma_{k}-m_{\text{adj}})}{\text{sin}(-\sigma_{j}\pm \sigma_{k}-m_{\text{adj}})}\prod_{a=1}^{N_{f}}\dfrac{\text{sin}(\sigma_{j}-m_{a})}{\text{sin}(-\sigma_{j}-m_{a})}=1
\end{equation*}
could correspond to the Bethe equation (\ref{25}). Obviously, we see that the representation has one more item, anti-fundamental representation, than (\ref{90}).

In this paper, we find that the vacuum equations from alternate effective superpotential, which also are suitable for the same integrable spin chains. With this expression, we give the corresponding duality between the vacuum equations and the Bethe equations. The following are the differences between \cite{DZ23a} and this work. 

For the $\text{A}$-type supersymmetry gauge theory, the two vacuum equations both need the condition $N_{f}=N_{f}^{'}$ and $m_{a}=m_{a}^{'}$ to match the same closed $\text{XXZ}$ spin chain, and the two dictionaries of the Bethe/Gauge correspondence are also the same. For the $\text{BCD}$-type supersymmetry gauge theory, we only need to consider the adjoint representation as well as the fundamental representation. In this work, here we peel off $N_{f}^{'}$ parts, only need to consider $N_{f}$ \cite{DZ23a}, the choice here is naive and concise. From Lie theory, the $\text{ADE}$ series are self-dual, the $\text{B}_N$ and $\text{C}_N$ are Langlands dual to each other. The Bethe equation for a given boundary parameter matches only one branch of the squared root of the vacuum equation. In this way, the two branches for the $\text{ADE}$ are the same. But for the $\text{B}_N$ and $\text{C}_N$ case, they belong to different branches. It should be noted that in the case of $\text{C}$-type gauge theory, we take the negative one in \cite{DZ23a}, so we cannot treat the $\text{B}_N$ and $\text{C}_N$ gauge theory with equally foot there.

Compared with the last work, the boundary parameters of the $\text{XXZ}$ spin chain which match the $\text{B}$-type gauge theory and the $\text{C}_{N}$-type gauge theory are different, while the boundary parameter of the $\text{D}_{N}$-type gauge theory are the same. In this article, to get the Bethe/Gauge correspondence, the spins of particle of relating spin chain are no restriction for $\text{C}_{N}$-type gauge theory and $\text{D}_{N}$-type gauge theory. Such that, we can think that for simple-laced Lie algebras, the duality obtained by these two kinds of effective superpotentials are the same, but for non-simple-laced Lie algebras, $\text{B}$-type and $\text{C}$-type gauge theory are Langlands duals under these two derivations.

In the case of the $\text{B}_{N}$-type gauge theory, to make the duality with spin chain explicitly, we have to set any two or three sites spins of the one dimension spin chain as $s_{a}=-1/2$. If the two boundary conditions $\xi$ are the same, we only need to fix two sites; otherwise, we need to fix three ones. If we regard the boundary as spin state with $s_{b}=1/2$, by the Pauli exclusion principle, the spin site fixed by a boundary will be reversed. If $s_{a}=-1/2$, the eigenvalues $s(s+1)=-1/4$. However, in physics, the value of spin for a particle cannot be negative.
Such that, In the even case, it needs to fix $2n$-site with $s_{a}=-1/2$ to make up the spin reversed  effect. In this way, 2-site with $s_{a}=-1/2$ is the simplest choice. 

By the way, in $\text{B}_{N}$ case with odd number spin sites fixed, which branch of Bethe's equation (BE) matches the vacuum equation, is depending on our choice of boundary parameters of the spin chain. In other word, the two branches of the vacuum equations are suitable. Also, the negative one of $\text{B}$-type gauge theory is dual to the positive one of $\text{C}$-type gauge theory.

On the other hand, in $\text{B}_{N}$ case with even number spin sites fixed, we have to choose the positive part of the vacuum equations to fit the Bethe equations. The result is exactly dual to the $\text{C}$-type gauge theory in \cite{DZ23a}. Equivalently speaking, the choice of the negative branch of the vacuum equation of the $\text{C}$-type gauge theory can be made up by the negative spin of the spin chain. We can regard the spin reversed effect is a result of the fixed boundary condition. Recall the hole effect in physics. If an electron is removed from an orbital that is normally occupied in the ground state, the analogy is made with the Dirac theory of positrons. A hole is a quasiparticle that has charge $+|e|$ with spin $1/2$ , and energies and velocities similar to quasi-electrons. In similarly to electron-hole effect, as a spin site fixed by a boundary, the spin of the site will be reversed. We call this as the boundary-spin effect. This is a very interesting phenomenon, and for our knowledge, there are no such duality is known in the literature to the present yet. We hope the boundary-spin effect can be observed and investigated in certain physics system.

\begin{figure}[htbp]
\centering
\includegraphics[width=.9\textwidth]{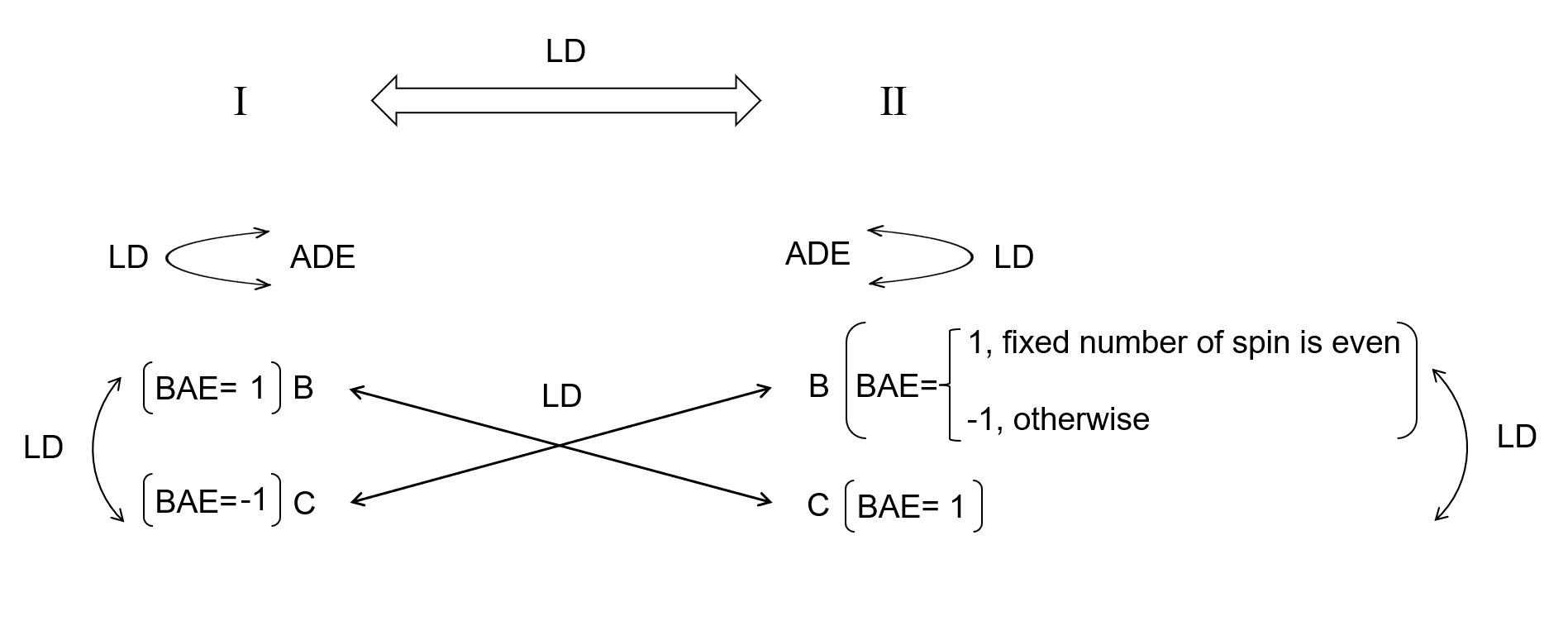}
\caption{The Langlands dualities between the work in \cite{DZ23a} (I) and the results here (II).\label{fig:i}}
\end{figure}

For the dilogarithm function, there is the factorization formula \cite{L81}
\begin{equation*}
\text{Li}_{2}(e^{rx})=r(\text{Li}_{2}(e^{x})+ \text{Li}_{2}(\omega e^{x})+\cdots+\text{Li}_{2}(\omega^{r-1} e^{x}))
\end{equation*}
where $\omega=e^{2\pi i/r}$, and $r$ depends only on the choice of the gauge groups in this article. Especially, 
\begin{equation*}
\text{Li}_{2}(e^{2x})=2(\text{Li}_{2}(e^{x})+ \text{Li}_{2}(-e^{x}))
\end{equation*}
This is the reason why there are more copies (anti-) fundamental representations needed 
in \cite{DZ23a}. From the Bethe/Gauge correspondence point of view, these two kinds of realizations can be viewed the effective superpotential (\ref{96}) in \cite{DZ23a} and the effective superpotential (\ref{7}) in this article as a Langlands duality. For different gauge groups, the effective superpotential (\ref{7}) and the previous effective superpotential (\ref{96}) are also the same relationship. The duality relationship is illustrated in figure \ref{fig:i}. The left side represents the work \cite{DZ23a} (I), while the right side represents the current work (II).

\section{Conclusion and discussion}\label{f}

In this paper, we give a novel Bethe/Gauge correspondence between classical Lie groups supersymmetry gauge theory and spin chains with closed boundary condition or open boundary condition. To be specific, we give the new effective superpotential depending on the roots of Lie algebras and get the vacuum equation corresponding to 
different Lie groups. Comparing with the results in \cite{DZ23a}, we can see that the derivation is easier, the duality pictures are more obvious, and the results are more concise. What is more, the parameters of the spin chain in this work also contain more information. We can view them as a new kind of realization of the Langlands dual relation between $\text{B}$-type gauge theory and $\text{C}$-type gauge theory. For simple-laced Lie algebras, the two types of different effective superpotential magically correspond to the same vacuum equations. We can regard the two kinds of realizations for the effective superpotential are the Langlands duality further.

There are several essays in which our results can be generalized. A direct extension is to lift the 3d supersymmetry gauge theory to 4d or higher dimensions cases \cite{CK18,NP23}. At the same time, it can be guessed that the open $\text{XXZ}$ spin chain should be upgraded to the $\text{XYZ}$ model \cite{FHSY96}. Another consideration is quiver gauge theory, which has richer content to study \cite{GK13,NW21}. Our results can also be generalized to the case of quiver gauge theory with the same product gauge groups or different product gauge groups.

\acknowledgments
The financial support from the Natural Science Foundation of China (NSFC, Grants 11775299) is gratefully acknowledged from one of the authors (Ding).

\appendix 

\section{Spin chain and Bethe ansatz equation}\label{A}

In this appendix, we give a brief introduction to $\text{XXX}$ spin chain and $\text{XXZ}$ spin chain for $SU(2)$. The integrability of a spin chain is characterized by an $R$-matrix, $R(u):V\otimes V \rightarrow V\otimes V$, satisfying the Yang-Baxter equation,
\begin{equation}\label{101}
R_{12}(u-v)R_{13}(u)R_{23}(v)=R_{23}(v)R_{13}(u)R_{12}(u-v)
\end{equation}
where $u,v$ are called the spectral parameters. The object $R_{ij}$ are linear operators in the tensor product of the three linear space $V\otimes V\otimes V$ with $R_{12}=R(u)\otimes I$, $R_{23}=I\otimes R(u)$, etc. The most general $R$-matrix for a solvable $\text{XXZ}$ spin chain model can be expressed as \cite{Bax07}
\begin{equation}\label{102}
R^{\text{XXZ}}(u)=\begin{pmatrix}
[u+\eta]&0&0&0\\
0&[u]&[\eta]&0\\
0&[\eta]&[u]&0\\
0&0&0&[u+\eta]
\end{pmatrix}
\end{equation}
where 
\begin{equation}\label{}
[x]:=\dfrac{\text{sin}(\pi x)}{\text{sin}(\pi \eta)}
\end{equation}
where $\eta$ is the crossing parameter. For $\text{XXZ}$ spin chain with periodic boundary condition, let the monodromy matrix be
\begin{equation}\label{103}
T_{0}(u)=R_{0L}(u-\vartheta_{L})\cdots R_{01}(u-\vartheta_{1}):=\begin{pmatrix}
A(u)&B(u)\\
C(u)&D(u)
\end{pmatrix}
\end{equation}
where $T(u)\in \text{End}(V^{(0)}\otimes V^{\otimes L})$, $V^{(0)}$ is an auxiliary space, $A(u),B(u),C(u),D(u)\in \text{End}(V^{\otimes L})$, and {$\{\vartheta_{1},\cdots, \vartheta_{L}\}$} are called inhomogeneous parameters. The monodromy matrix satisfies the $\text{RTT}$-relation, 
\begin{equation*}\label{}
R_{12}(u-v)T_{1}(u)T_{2}(v)=T_{2}(v)T_{1}(u)R_{12}(u-v)
\end{equation*}
which can be proven by substituting the definition of $T_{0}(u)$ in (\ref{103}) and using the Yang-Baxter equation (\ref{101}). The transfer matrix is given by
\begin{equation}\label{104}
t(u)=\text{tr}_{0}T_{0}(u)=A(u)+D(u)
\end{equation}
From the $\text{RTT}$-relation, the transfer matrices
corresponding to different spectral parameters commute \cite{Fad96}:
\begin{equation}
[t(u),t(v)]=0
\end{equation}
which ensures the integrability of this system. For a given ground state $\Omega$ of the system, the fact that
\begin{equation}\label{105}
\prod_{i=1}^{M}B(u_{i})\Omega
\end{equation}
is the Bethe ansatz state, which is motivated by the commutation relations of $[B,B]=0$ from the $\text{RTT}$ equation, with $M$ is the number of magnons in our spin chain and $u_{i}$ is the Bethe roots solution of the Bethe equation. Using the transfer matrix (\ref{104}) to act on the Bethe ansatz state (\ref{105}), and the commutation relation between $A(u)$, $D(u)$ and $B(u)$ from the $\text{RTT}$ relation, we get the Bethe ansatz equation with periodic boundary condition
\begin{equation}\label{106}
\prod_{a=1}^{L}\dfrac{[u_{i}+\frac{\eta}{2}+\eta s_{a}-\vartheta_{a}]}{[u_{i}+\frac{\eta}{2}-\eta s_{a}-\vartheta_{a}]}=\prod_{j\neq i,j=1}^{M}\dfrac{[u_{i}-u_{j}+\eta]}{[u_{i}-u_{j}-\eta]}
\end{equation}
where we take spin $s_{a}$ representation at the $a$-th site. 

For open $\text{XXZ}$ spin chain, we focus on the open $\text{XXZ}$ spin chains with diagonal boundary operator $K$-matrix
\begin{equation*}\label{}
K^{\text{XXZ}}(u,\xi)=\begin{pmatrix}
[u+\xi]&0\\
0&-[u-\xi]
\end{pmatrix}
\end{equation*}
The transfer matrix of an open chain is given \cite{WYC15}
\begin{equation}\label{107}
t(u)=\text{Tr}_{0}K_{+}(u)T_{0}(u)K_{-}(u)T_{0}^{-1}(-u)
\end{equation} 
where {$T(u)$ is usually taken to be the same one in closed spin chain, $K_{-}(u)\in \text{End}(V^{(0)})$ and $K_{+}(u)\in \text{End}(V^{(0)})$ are the boundary operators}, which satisfy the Reflection Equation (RE) and dual RE, respectively,
\begin{equation*}\label{}
\begin{aligned}
&R_{12}(u-v)K_{-}^{1}(u)R_{21}(u+v)K_{-}^{2}(v)\\
&=K_{-}^{2}(v)R_{21}(u+v)K_{-}^{1}(u)R_{12}(u-v)\\
\end{aligned}
\end{equation*}
and
\begin{equation*}
\begin{aligned}
&R_{12}(-u+v)K_{+}^{1}(u)R_{21}(-u-v-2\eta)K_{+}^{2}(v)\\
&=K_{+}^{2}(v)R_{21}(-u-v-2\eta)K_{+}^{1}(u)R_{12}(-u+v)
\end{aligned}
\end{equation*}
where $K_{-}^{1}(u):=K_{-}(u)\otimes id_{V_{2}}$ and $K_{-}^{2}(u):=id_{V_{1}} \otimes K_{-}(u)$. Rather than the one-row monodromy matrix, we need to introduce the double-row monodromy matrix 
\begin{equation}\label{108}
U_{-}(u)=T(u)K(u-\frac{\eta}{2},\xi_{-})\sigma_{y}T^{t}(-u)\sigma_{y}:=\begin{pmatrix}
\mathcal{A}(u)&\mathcal{B}(u)\\
\mathcal{C}(u)&\mathcal{D}(u)
\end{pmatrix}
\end{equation} 
The equivalent of the $\text{RTT}$ relation in case of open spin chain is the $\text{RURU}$ relation.
\begin{equation}\label{}
\begin{aligned}
&R_{12}(u-v)U_{-}^{1}(u)R_{12}(u+v-\eta)U_{-}^{2}(v)\\
&=U_{-}^{2}(v)R_{12}(u+v-\eta)U_{-}^{1}(u)R_{12}(u-v)
\end{aligned}
\end{equation}
Just like the closed spin chain, the ground state $\Omega$ is also the ground state of an open chain with diagonal boundary condition. The Bethe ansatz state is 
\begin{equation*}\label{}
\prod_{i=1}^{M}\mathcal{B}(u_{i})\Omega
\end{equation*}
Define $\tilde{\mathcal{D}}(u)=[2u]\mathcal{D}(u)-[\eta]\mathcal{A}(u)$. For the $\text{XXZ}$ spin chain with diagonal boundary condition, the transfer matrix $t(u)$ is constructed by the double-row monodromy matrix as
\begin{equation*}\label{}
\begin{aligned}
t(u)&=\text{Tr}\left(K_{+}(u+\eta,\xi_{+})U_{-}(u)\right)\\
&=\dfrac{[2u+\eta][u-\frac{\eta}{2}+\xi_{+}]}{[2u]}\mathcal{A}(u)-\dfrac{[u+\frac{\eta}{2}-\xi_{+}]}{[2u]}\tilde{\mathcal{D}}(u)
\end{aligned}
\end{equation*}
where
\begin{equation*}
 K_{+}(u,\xi_{+})=K(u+\frac{\eta}{2},\xi_{+}):=\begin{pmatrix}
 [u+\frac{\eta}{2}+\xi_{+}]&0\\
 0&-[u+\frac{\eta}{2}-\xi_{+}]
\end{pmatrix}
\end{equation*}
The Bethe equation of the open $\text{XXZ}$ spin chain with diagonal boundary condition is given by
$$t(u)\prod_{i=1}^{M}\mathcal{B}(u_{i})\Omega$$
Finally, the simplified equation is
\begin{equation}\label{109}
\begin{aligned}
&\dfrac{\text{sin}[\pi (u_{i}-\frac{\eta}{2}+\xi_{+})]}{\text{sin}[\pi(u_{i}+\frac{\eta}{2}-\xi_{+})]}\dfrac{\text{sin}[\pi (u_{i}-\frac{\eta}{2}+\xi_{-})]}{\text{sin}[\pi(u_{i}+\frac{\eta}{2}-\xi_{-})]}\\
&\times\prod_{a=1}^{L}\dfrac{\text{sin}[\pi (u_{i}+\frac{\eta}{2}+\eta s_{a}-\vartheta_{a})]\text{sin}[\pi(-u_{i}+\frac{\eta}{2}-\eta s_{a}-\vartheta_{a})]}{\text{sin}[\pi(-u_{i}+\frac{\eta}{2}+\eta s_{a}-\vartheta_{a})]\text{sin}[\pi(u_{i}+\frac{\eta}{2}-\eta s_{a}-\vartheta_{a})]}\\
&\times \prod_{j\neq i,j=1}^{M}\dfrac{\text{sin}[\pi (u_{j}+u_{i}-\eta)]\text{sin}[\pi(u_{j}-u_{i}-\eta)])}{\text{sin}[\pi(u_{j}-u_{i}+\eta)]\text{sin}[\pi(u_{j}+u_{i}+\eta)]}=1
\end{aligned}
\end{equation}

The Bethe ansatz equations of $\text{XXX}$ spin chain can be obtained by degeneration of (\ref{19})
\begin{equation}\label{110}
\begin{aligned}
&\dfrac{(u_{j}-\frac{\eta}{2}+\xi_{+})}{(u_{j}+\frac{\eta}{2}-\xi_{+})}\dfrac{(u_{j}-\frac{\eta}{2}+\xi_{-})}{(u_{j}+\frac{\eta}{2}-\xi_{-})}\prod_{a=1}^{L}\dfrac{(u_{j}+\frac{\eta}{2}+\eta s_{a}-\vartheta_{a})(-u_{j}+\frac{\eta}{2}-\eta s_{a}-\vartheta_{a})}{(-u_{j}+\frac{\eta}{2}+\eta s_{a}-\vartheta_{a})(u_{j}+\frac{\eta}{2}-\eta s_{a}-\vartheta_{a})}\\
&\times \prod_{k\neq j,k=1}^{M}\dfrac{ (u_{k}-u_{j}-\eta)(u_{k}+u_{j}-\eta)}{(u_{k}-u_{j}+\eta)(u_{k}+u_{j}+\eta)}=1
\end{aligned}
\end{equation}



\bibliographystyle{plain}

\begin{thebibliography}{abcdefghijk}

\bibitem[Bax07]{Bax07}
R. Baxter, \emph{Exactly Solved Models in Statistical Mechanics}, Dover books on physics, Dover Publications(2007). https://books.google.ie/books?id=G3owDULfBuEC.

\bibitem[CK18]{CK18}
Heng-Yu Chen and Taro Kimura, \emph{Quantum integrability from non-simply laced quiver gauge theory}, \emph{JHEP} {\bf 16} (2018) 165 [arXiv:1805.01308].

\bibitem[DZ23a]{DZ23a}
Xiang-Mao Ding and Tinglyer Zhang, \emph{Bethe/Gauge Correspondence for \text{ABCDEFG}-type 3d Gauge Theories}, \emph{JHEP} {\bf 04} (2023) 036 [arXiv:2303.03102].

\bibitem[DZ23b]{DZ23b}
Xiang-Mao Ding and Tinglyer Zhang, \emph{Bethe/gauge correspondence for linear quiver theories with \text{ABCD} gauge symmetry
and spin chains}, \emph{Nuclear Physics B} {\bf 991} (2023) 116222 [arXiv:2023.04575]

\bibitem[Fad96]{Fad96}
L. Faddeev, \emph{How algebraic Bethe ansatz works for integrable model}, [arXiv:hep-th/9605187].

\bibitem[FHSY96]{FHSY96}
H. Fan, B.-Y. Hou, K.-J. Shi and Z.-X. Yang, \emph{Algebraic Bethe ansatz for eight vertex model with general open boundary conditions}, \emph{Nucl. Phys. B} {\bf478} (1996) 723 [hep-th/9604016].
[INSPIRE].

\bibitem[GK13]{GK13}
D. Gaiotto and P. Koroteev, \emph{On three dimensional quiver gauge theories and integrability}, \emph{JHEP} {\bf 05} (2013) 126.

\bibitem[GNO77]{GNO77}
P. Goddard, J. Nuyts, D. Olive, \emph{Gauge theories and magnetic charge}, Nucl. Phys. B 125 (1977) 1.

\bibitem[GKMMM95]{GKMMM95}
A. Gorsky, I. Krichever, A. Marshakov, A. Mironov, A. Morozov, \emph{Integrability and Seiberg-Witten exact solution}, \emph{Phys. Lett. B} {\bf355} (1995) 466 [hep-th/9505035].

\bibitem[Hum72]{Hum72}
J.Humphreys, \emph{Introduction to Lie Algebra and Representation Theory}, Springer Verlag, New York. http://dx.doi.org 10.1007/978-1-4612-6398-2.

\bibitem[KZ21]{KZ21}
T. Kimura and R.-D Zhu, \emph{Bethe/gauge correspondence for SO/Sp gauge theories and open spin chains}, \emph{JHEP} {\bf 03} (2021) 227 [arxiv:2012.14197].

\bibitem[LN21]{LN21}
N. Lee and N. Nekrasov, \emph{Quantum spin systems and supersymmetric gauge theories. Part I}, \emph{JHEP} {\bf 03} (2021) 093 [arXiv:2009.11199].


\bibitem[L81]{L81}
L. Lewin, \emph{ Polylogarithms and Associated Functions}, Elsevier North Holland 1981.

\bibitem[NP23]{NP23}
N. Nekrasov and V. Pestun, \emph{Seiberg-witten geometry of four-dimensional n = 2 quiver gauge theories},
\emph{SIGMA} {\bf 19} (2023), 047  [arXiv:1211.2240].

\bibitem[NPS18]{NPS18}
N. Nekrasov, V. Pestun and S.L. Shatashvili, \emph{Quantum Geometry and Quiver Gauge Theories}, \emph{Commun. Math. Phys}. {\bf 357}, (2018) 519.

\bibitem[NS09a]{NS09a}
N.Nekrasov and S.L. Shatashvili, \emph{Supersymmetric vacua and Bethe ansatz}, \emph{Nucl. Phys. B Proc. Suppl.} {\bf192–193} (2009) 91 [arXiv:0901.4744] [INSPIRE].

\bibitem[NS09b]{NS09b}
N. Nekrasov and S.L. Shatashvili, \emph{Quantum integrability and supersymmetric vacua}, \emph{Prog. Theor. Phys. Suppl.} {\bf177} (2009) 105 [arXiv:0901.4748] [INSPIRE].

\bibitem[NS09c]{NS09c}
N.A. Nekrasov and S.L. Shatashvili, \emph{Quantization of Integrable Systems and Four Dimensional Gauge Theories,} in \emph{16th International Congress on Mathematical Physics}, (2009), DOI [arXiv:0908.4052] [INSPIRE].

\bibitem[NW21]{NW21}
H. Nakajima and A. Weekes, \emph{Coulomb branches of quiver gauge theories with symmetrizers}, \emph{J. Eur. Math. Soc.}, {\bf 25} (2023) 203.

\bibitem[SW94a]{SW94a}
N. Seiberg and E. Witten, \emph{Electric-magnetic duality, monopole condensation, and confinement in N = 2 supersymmetric Yang-Mills theory}, \emph{Nucl. Phys. B} {\bf426} (1994) 19 [\emph{Erratum ibid}. {\bf430} (1994) 485] [hep-th/9407087] [INSPIRE].

\bibitem[SW94b]{SW94b}
N. Seiberg and E. Witten, \emph{Monopoles, duality and chiral symmetry breaking in N = 2 supersymmetric QCD}, \emph{Nucl. Phys. B} {\bf431} (1994) 484 [hep-th/9408099] [INSPIRE].

\bibitem[S95]{S95}
N. Seiberg, Electric–magnetic duality in supersymmetric non Abelian gauge theories, Nucl. Phys. B 435 (1995)129, arXiv:hep-th/9411149

\bibitem[SW97]{SW97}
N. Seiberg and E. Witten, \emph{Gauge dynamics and compactification to three-dimensions, in The mathematical beauty of physics: A memorial volume for Claude Itzykson, C. Itzykson et al. Gauge dynamics and compactification to three-dimensions, in The mathematical beauty of physics: A memorial volume for Claude Itzykson, C. Itzykson et al.}, \emph{World Scientific, Singapore} (1997) [arXiv:9607163] [INSPIRE].

\bibitem[WYC15]{WYC15}
Y. Wang, W. L. Yang, J. Cho et al, \emph{Off-diagonal Bethe Ansatz for exactly solvable models[M]}, Springer Berlin Heidelberg(2015).

\bibitem[YS20]{YS20}
Y. Yoshida and K. Sugiyama, \emph{Localization of three-dimensional $\mathcal{N}$=2 supersymmetric theories on $S^{1} \times D^{2}$}, \emph{PTEP} {\bf2020} (2020) 113B02 [arXiv:1409.6713] [INSPIRE].

\bibitem[YY15]{YY15}
M. Yamazaki and W. Yan, \emph{Integrability from 2d $\mathcal{N}=(2,~2)$dualities}, 
\emph{J. Phys. A: Math. Theor. {\bf 48} (2015) 394001} [arxiv:1504.05540]




\end{thebibliography}

\end{document}